\documentclass[twocolumn,showpacs,showkeys,preprintnumbers,amsmath,amssymb]{revtex4}

\usepackage{graphicx}
\usepackage{dcolumn}
\usepackage{bm}

\begin{document}

\preprint{FT-HEP001}

\title{About the inclusion of an infinite number of resonances in anomalous decays.}

\author{D. Garc\'{\i}a Gudi\~no and G. Toledo S\' anchez}
\affiliation{Instituto de F\'{\i}sica,  Universidad Nacional Aut\'onoma de M\'exico, AP 20-364,  M\'exico D.F. 01000, M\'exico}%

\date{\today}

\begin{abstract}
 The usual procedure to extract the $g^{eff}_{\omega\rho\pi}$ effective coupling from experimental data resumes not only the $\rho$ meson effect but also all its additional radial excitation modes. In this work, we consider a particular form of the spectrum and relations among the couplings to explicitly add the radial excitations of the $\rho$ meson from the $\omega \rightarrow \pi^0 \gamma$ and $\omega \rightarrow 3\pi$ decays. This allows to identify the single  $g_{\omega\rho\pi}= 8.3 \pm 0.8  \ {\text GeV}^{-1}$, which is about 40\% smaller than the effective  $g^{eff}_{\omega\rho\pi}$.
We verify the consistency with the chiral approach in the  $\pi^0 \rightarrow \gamma\gamma$ and $\gamma^* \rightarrow 3\pi$ processes, exhibiting the role of each contribution. In particular, we show that for the $\gamma^* \rightarrow 3 \pi$ decay, the usual relation $\mathcal{A}^{VMD}_{\gamma3\pi}=(3/2)\mathcal{A}^{WZW}_{\gamma3\pi}$, encodes all the vector contributions and not only the $\rho$ meson one. In addition, we find that there is a strong (accidental) cancelation between the radial excitations and the contact term contributions.
\end{abstract}

\maketitle

\section{Introduction}
The study of the low energy regime of the strong interaction among quarks relies on the use of effective theories. A description based on the effective hadronic degrees of freedom, namely the vector meson dominance approach (VMD), assigns an effective coupling to the hadronic interaction, which must be determined from the experimental information.
Relations between these parameters and those coming from low energy theorems  \cite{lowetheorem1,lowetheorem2,lowetheorem3,anomaly1,anomaly2} can be drawn by matching both descriptions \cite{ksfr} and building models which incorporate vector mesons into the chiral symmetric lagrangians \cite{kura,kay,fujiwara,kura1,xral,xral2,prades94}.
In particular, experimental information from several observables is currently available to determine the coupling between the $\omega$, $\rho$ and $\pi$ mesons \cite{cmd200,cmd2new,kloe,aulchenko,achasov02,achasov03,pdg}.
 Estimates based on different theoretical approaches \cite{lutz09,Khathimovsky85,Margvelashvili88,sumrules1,sumrules2,sumrules3,lublinsky97,cesareo,su3a,su3b,su3break1,su3break2,khun06,lucio,maris03} find it to be in the range  from 7.3 to 17 GeV$^{-1}$. Based on the VMD approach, we have recently analyzed the data and found that such coupling lies in the range from 11.9 to 15.7 GeV$^{-1}$ \cite{davidIJMPA}, all these points out to possible experimental and/or theoretical problems.\\ 
The experimental issues, if any, will be settled with the advent of more precise measurements. On the other hand, the theoretical treatment considers that, since the phase space forbids the $\omega$ to decay into the $\rho$ and $\pi$ mesons, at least one of them must be off-shell. To deal with the off-shell features in the VMD approach, a systematic inclusion of the infinite number of excitations of the $\rho$ ground state must be included. However, it is common, in a first approach, to neglect them all. That is, the extracted effective strong coupling, labeled as $g^{eff}_{\omega\rho\pi}$, resumes all possible additional contributions from higher  states and not only the $\omega-\rho-\pi$ interaction.
A proper description to include all of them would require the knowledge of the full excited spectrum and their corresponding couplings. A difficult task to accomplish, since there is no information on the total spectrum. Attempts in this direction have been made by invoking the factorizable dual model \cite{cesareo}, which considers the vector meson spectrum to follow the form $m^2_{\rho_n}=m^2_{\rho}(1+2n)$, and the ratio of the coupling constants  $g_{\rho_n\pi\pi}/g_{\rho_n} $ is determined by requiring  the form factor accounting for all the off-shell effects to be a ratio of gamma functions, fitted to the pion form factor. As a result, all the tree level diagrams are modified by such form factor at each  vertex containing an off-mass-shell meson. The consistency of that description with the chiral anomaly for $\pi \to \gamma\gamma$, gives $g_{\omega\rho\pi}=-16\pm 1$ GeV$^{-1}$.\\ 
In this work, we estimate the contribution from the $\rho$ meson radial excitations to the $g^{eff}_{\omega\rho\pi}$ coupling. We consider the spectrum to be given by $m_{\rho_n}^2=m_\rho^2(1+a n)^2$, (a=constant) as inspired by the image of mesons built up of  a quark and an antiquark bound through a harmonic oscillator potential  \cite{chen-tsai}. We also assume a KSFR-like relation \cite{ksfr} for each $\rho_n\pi\pi$ vertex and SU(3) symmetry (see section III). 
 
 The article is organized as follows: In section II we review  the  $\pi^0 \rightarrow \gamma\gamma$ and $\gamma^* \rightarrow 3\pi$ processes to show the standard procedure to describe them in the VMD and the chiral approach, and the conditions to match such descriptions. We also point out where the radial excitations contribution is being hidden. Then, the $\omega \rightarrow 3\pi$ is used to illustrate the changes the $g^{eff}_{\omega\rho\pi}$ can have when invoking the matching conditions.
 In section III we establish the assumptions  regarding the radial excitations spectrum and couplings. Then, we proceed to determine the effect of the $\rho$ radial excitations in the $g^{eff}_{\omega\rho\pi}$ coupling using the $\omega \rightarrow \pi^0 \gamma$ and $\omega \rightarrow 3\pi$ decays, which are free of $\omega$ radial excitations. We verify the consistency with the chiral approach in the  $\pi^0 \rightarrow \gamma\gamma$ and $\gamma^* \rightarrow 3\pi$ processes and its implications. In section IV we discuss our results.

\section{Baseline}
Let us  make a review of the results from the decays when no radial excitations are included. This will be the baseline of the calculation to compare with.\\
The VMD Lagrangian including the $\rho$, $\pi$ and $\omega$ mesons can be set as:
\begin{eqnarray}
{\cal L}&=& g_{\rho\pi\pi} \epsilon_{abc} \rho_\mu^a \pi^b \partial^\mu \pi^c 
+g_{\omega\rho\pi}\delta_{ab}\epsilon^{\mu\nu\lambda\sigma}\partial_\mu \omega_\nu \partial_\lambda \rho_\sigma^a  \pi^b \nonumber \\
&+&g_{3\pi} \epsilon_{abc}   \epsilon^{\mu\nu\lambda\sigma}\omega_\mu \partial_\nu \pi^a  \partial_\lambda  \pi^b  \partial_\sigma \pi^c +
\frac{e m_V^2}{g_V}V_\mu A^\mu + ...
\label{lagrangian}
\end{eqnarray}
This Lagrangian exhibits only the relevant pieces for this work and should be part of any effective Lagrangian describing these mesons. Terms with higher
derivatives and additional terms which allow to preserve gauge invariance are not shown \cite{klz}. We have made explicit the notation regarding the couplings and the corresponding fields and, in the last term, $V$ refers in general to vector mesons and $A^\mu$ refers to the photon field. Here $g_V=2 \alpha \sqrt{\pi m_V/3\Gamma_{V \rightarrow l^+ l^-}}$ ($l=e,\mu,\tau$).\\
On the other hand, the chiral symmetry of the strong interaction dictates that, at the lowest order, the Wess Zumino Witten anomaly \cite{wzw} is responsible for the $\pi^0 \rightarrow \gamma \gamma$ and $\gamma^* \rightarrow 3\pi$ decays.  For our purposes the relevant part of the Lagrangian is given by:
\begin{eqnarray}
{\cal L}^{WZW}&=&\frac{i N_C e^2}{24 \pi^2 f_\pi}   \epsilon^{\mu\nu\lambda\sigma}\pi^0 F_{\mu\nu}F_{\lambda \sigma}\\
&+&\frac{N_C}{3\times6}\frac{e}{4 \pi^2 f^3_\pi}
\epsilon_{abc}   \epsilon^{\mu\nu\lambda\sigma}
B_\mu \partial_\nu \pi^a  \partial_\lambda  \pi^b  \partial_\sigma \pi^c +...,\nonumber
\end{eqnarray}
where $N_c$ is the number of colors and $f_\pi=0.093$ GeV, and we did split  the coefficients for the sake of clarity when writing the amplitudes below.

\subsection{The $\pi^0 \rightarrow \gamma\gamma$ decay}
The amplitude for the process has a generic form:
\begin{equation}
{\cal M}_{\pi\gamma\gamma}= \epsilon_{\mu\nu\lambda\sigma} k^\mu_1 \eta^\nu_1 k^\lambda_2 \eta^\sigma_2 {\mathcal A}_{\pi\gamma\gamma},
\end{equation}
where $k_1$ ($\eta_1$) and $k_2$ ($\eta_2$) are the photons momenta (polarizations) and ${\mathcal A}_{\pi\gamma\gamma}$ encodes the model details. In the chiral approach it is given by the WZW, as mentioned above, while in the VMD approach, the decay can be seen as proceeding through the emission of the $\rho$ and $\omega$ mesons which eventually decay into photons.
Using the above Lagrangians, and taking the zero momentum limit, they correspond to:
\begin{equation}
{\mathcal A}^{VMD}_{\pi\gamma\gamma}=\frac{2e^2 g_{\omega\rho\pi}}{g_\rho g_\omega};
\hspace*{1.5cm}
{\mathcal A}^{WZW}_{\pi\gamma\gamma}=\frac{6 e^2}{24\pi^2 f_\pi}.
\label{piggamplitudes}
\end{equation}
\noindent Note that ${\mathcal A}^{WZW}_{\pi\gamma\gamma}=0.025 $ GeV$^{-1}$ for $f_\pi=0.93$ MeV, to be compared with the experimental value  ${\mathcal A}_{\pi\gamma\gamma}=0.025 \pm 0.001$ GeV$^{-1}$. \cite{pdg}. Thus, the anomaly completely describes the process and, at this level,  there is no need to include higher order corrections which should incorporate vector mesons in a chiral invariant way.\\
Although both WZW and VMD descriptions are different, we can {\em match both amplitudes} to relate the  couplings. Namely, 
\begin{equation}
|g^{eff}_{\omega\rho\pi}| = |g_{\rho\pi\pi}g_{\omega}/8\pi^2f_\pi |.
\label{c1}
\end{equation}
Note that this parameter, initially considered as due to a single channel ($\rho$ and $\omega$), when imposed to account for the total effect of the anomaly becomes an {\em effective coupling resuming all the possible contributions in the VMD framework}. This observation will be crucial in the interpretation of the corresponding magnitude.\\
An additional relationship between the $\rho$ and the pion properties is given by the so-called KSFR relation\cite{ksfr}:
\begin{equation}
 g_{\rho\pi\pi}=\frac{m_\rho}{\sqrt{2} f_\pi}. 
 \end{equation}
Considering that the KSFR relation, SU(3) symmetry ($g_\omega=3g_\rho$) and universality condition ($g_\rho=g_{\rho\pi\pi}$) are hold, the effective coupling Eqn. (\ref{c1}) becomes
\begin{equation}
 |g^{eff}_{\omega\rho\pi}|=\frac{ 3m_\rho^2}{16 \pi^2f_\pi^3}=14.2 \ {\text GeV}^{-1}.
 \label{e1gorp}
\end{equation}

\subsection{The $\gamma^* \rightarrow 3\pi$ decay}
 The amplitude of the process can be written in general as:
 \begin{equation}
\mathcal{M} = \imath \epsilon _{\mu \alpha \beta \gamma} \eta^{\mu} p_1^{\alpha }p_2^{\beta }p_3^{\gamma } \mathcal{A}_{\gamma3\pi},
\label{ampw3pi}
\end{equation} 
where $\mathcal{A}_{\gamma3\pi}$ encodes the details of the model used to describe the process,  and $p_1$, $p_2$, $p_3$ are the pions momenta and $\eta$ is the photon polarization.

The decay of the photon into  three pions in the chiral description also has its origin in the WZW anomaly, as discussed above. In the VMD approach, it has been shown that this decay is mainly produced through the $\omega$ into $\rho \pi$ decay channel \cite{gellman}, followed by the break down of the $\rho$ into another two pions.
At zero momentum they correspond to:
\begin{equation}
   \mathcal{A}^{WZW}_{\gamma3\pi}=\frac{e}{4\pi^2 f_\pi^3} \hspace*{1cm}
    \mathcal{A}^{VMD}_{\gamma3\pi}=\frac{6e}{g_\omega}\frac{ g^{eff}_{\omega \rho \pi}g_{\rho \pi \pi}}{m_\rho^2}
  \end{equation}
(a factor of 6 arises from the momenta permutations when bringing the amplitude to Eqn. \ref{ampw3pi} form)
Note that  $\mathcal{A}^{WZW}_{\gamma3\pi}=9.7$ GeV$^{-3}$, deviates from the experimental value of $\mathcal{A}_{\gamma3\pi}=12.9 \pm 0.9\pm 0.5$ GeV$^{-3}$, arguably by the momentum dependence effect at the measurement \cite{holstein}. 
Here we restrict ourselves to the zero momentum expression. By linking the anomaly term to the corresponding amplitude from VMD, the KSFR relation is not hold anymore. The reason lies on the fact that, although the $\rho$ channel approach is  well motivated, it is unable to fully capture the anomaly information. i.e. the decay have additional  axial contributions which can not be captured in the effective vector channel. Thus, in the VMD description, an additional  contact interaction between the $\omega$ and the 3$\pi$ mesons must be added:
\begin{equation}
{\cal L}^{c}_{\omega\pi\pi\pi}=g^c_{3\pi} \epsilon_{abc}   \epsilon^{\mu\nu\lambda\sigma}\omega_\mu \partial_\nu \pi^a  \partial_\lambda  \pi^b  \partial_\sigma \pi^c  
\end{equation}
where $g^c_{3\pi}$ is the corresponding effective coupling strength.
Then, by considering that the KSFR relation is hold and SU(3) symmetry ($g_\omega=3g_\rho$), the $\rho$ meson channel amplitude at zero momentum becomes:
\begin{equation}
  \mathcal{A}^{VMD(\rho)}_{\gamma 3\pi}=\frac{6e}{g_\omega}\frac{ g^{eff}_{\omega \rho \pi}g_{\rho \pi \pi}}{m_\rho^2}= \frac{3}{2}\frac{e}{4 \pi^2 f_\pi^3}=\frac{3}{2} \mathcal{A}^{WZW}_{\gamma3\pi},
\end{equation}
 that is three halves of the total amplitude from the Chiral anomaly \cite{wzw}. 
 Therefore, the contact term must account for the one half excess of the total amplitude. This was found by Rudaz \cite{rudaz} as a consistency requirement and by Cohen \cite{cohen} as the one which satisfies axial Ward identities.
  \begin{equation}
  \mathcal{A}^{c}= \frac{6e}{g_\omega}g^c_{3\pi} =\frac{-1}{2} \mathcal{A}^{WZW}_{\gamma3\pi}.
  \end{equation}
 
 \noindent This condition fixes the corresponding coupling to be: 
 \begin{equation}
 g^c_{3 \pi}=-\frac{g_{\rho\pi\pi}}{16\pi^2f_\pi^3}=- 47 \ GeV^{-3} ,\label{lowrelation}
\end{equation}
\noindent where we have made use of the relationship among the couplings as discussed above.

\subsection{The $\omega \rightarrow 3\pi$ decay}

This decay can be seen as a subprocess of the $\gamma^* \rightarrow 3\pi$ decay, and therefore the conclusions from the previous section can be extended to this case. Just let us point out that the contribution to the amplitude from the $\rho$ channel (using the same notation but  now $\eta$ is the $\omega$ polarization tensor):
\begin{equation}
\mathcal{M}_D = \imath \epsilon _{\mu \alpha \beta \gamma} \eta^{\mu} p_1^{\alpha }p_2^{\beta }p_3^{\gamma } \mathcal{A}_{\omega3\pi},
\label{ampf3pi}
\end{equation} 
where
\begin{equation}
\mathcal{A}_{\omega3\pi} =  \frac{6}{m_\rho^2}g^{eff}_{\omega \rho \pi } g_{\rho \pi \pi },
\label{Apdem}
\end{equation}
would require  $g^{eff}_{\omega \rho \pi } =15.7$ GeV$^{-1}$,
in order to  reach the 100\% of the experimental width \cite{davidIJMPA}. If we enforce the KSFR relation and add the contact term, the coupling becomes $g^{eff}_{\omega \rho \pi } =12.8$ GeV$^{-1}$.\\

The previous analysis relies mainly on the consideration that the vector channel is saturated by the $\rho$ meson and, upon matching VMD with the WZW anomaly, the $g^{eff}_{\omega \rho \pi}$ coupling constant is an effective coupling which not only accounts for the $\omega-\rho-\pi$ interaction but also all additional terms.
In order to determine the truly $g_{\omega \rho \pi}$ coupling,  the radial excitations spectrum and the corresponding couplings are required.  In the following we study a particular  model to include the radial excitations of the $\rho$ meson and show their implications.  

\section{Adding the radial excitations contribution.}
To include the radial excitations of the $\rho$ meson we make the following assumptions:\\
i) The spectrum of the radial excitations is given by: $m_{\rho_n}^2=m_\rho^2(1+a n)^2$  with $n=0,1,2,...$. This is the expected behavior for the quark and antiquark bound through a harmonic oscillator potential \cite{chen-tsai}. By construction, $n=0$ corresponds to the $\rho$ mass and for $n=1$ it is the $\rho'(1450)$, this fixes $a$=0.89.\\
ii) KSFR-like relation for each $\rho_n\pi\pi$ vertex, that is $g_{\rho_n}\equiv g_{\rho_n\pi\pi}=m_{\rho_n}/(\sqrt{2}f_\pi)$. This might be expected to follow from the fact that the effective lagrangian describing the radial excitations coupling to the pions have the same structure that the corresponding to the $\rho$ meson.\\

iii) SU(3) symmetry, which allows to relate $g_\omega=3g_\rho$.\\
iv) Under the above assumptions, the following relation among coupling constants is expected to hold $(g_{\omega \rho \pi}g_\rho)/(g_{\omega \rho_n \pi}g_{\rho_n})=1$, provided that, on dimensional grounds, $g_{\omega V \pi} \approx 1/m_V $. This dependence is cancelled out by the corresponding mass dependence on $g_V$ (assumption ii). Thus, in the ratio, the proportionality factor is expected to be of the same order in each case, since it accounts for the $\omega$ and $\pi$ mesons.\\

Note that, if we wanted to include the $\omega$ radial excitation on the same grounds, the above assumption ii) is no longer valid. Thus, we restrict ourselves only to the $\rho$ excitations.

\subsection{The $\omega \rightarrow \pi^0 \gamma$ decay}
As a first case, we will consider the $\omega \rightarrow \pi^0 \gamma$ decay to identify the  $g_{\omega \rho \pi}$ and the $\rho$ radial excitations effect. This process is clean in the sense that it is sensitive to the $\rho$ radial excitations but not to the $\omega$ ones. This fact allows to avoid any assumption on the $\omega$ excitations.

The amplitude for this process has the general form:
\begin{equation}
{\cal M}_{\omega\pi\gamma}=  i\epsilon_{\mu\nu\lambda\sigma} q^\mu_1 \eta^\nu k^\lambda_2 \epsilon^\sigma  \mathcal{A}_{\omega\pi\gamma},
\end{equation}
In this notation $ \mathcal{A}_{\omega\pi\gamma}$ encodes the model details. For this process, the VMD description, including the radial excitations, requires that at zero momentum:
\begin{equation}
 \mathcal{A}_{\omega\pi\gamma}= e\sum_{\rho_n}  \frac{g_{\omega \rho_n \pi}}{g_{\rho_n}}.  
\end{equation}

Now, using the above assumptions the global coupling can be set as:
\begin{eqnarray}
\sum_{\rho_n}   \frac{g_{\omega {\rho_n} \pi}}{g_{\rho_n}}&=& \sum_{\rho_n} \frac{g_{\omega \rho \pi}g_\rho}{g_{\rho_n}^2} = \sum_{\rho_n} \frac{g_{\omega \rho \pi}g_\rho  2f_\pi^2}{m_{\rho_n}^2} \nonumber\\
&=&
  \frac{g_{\omega \rho \pi}g_\rho 2f_\pi^2}{m_\rho^2 }\sum_{n=0}^\infty  \frac{1}{(1+an)^2} \nonumber\\
  &=&  \frac{g_{\omega \rho \pi}g_\rho 2f_\pi^2}{m_\rho^2 }\zeta(2,1/a)/a^2
\end{eqnarray}
where the convergence on the series above is equal to $\zeta(2,1/a)/a^2\approx 7/4$, with $\zeta$ the generalized Riemann Zeta function. Using  this in the amplitude, we can compute the decay width 
\begin{equation}
\Gamma(\omega \rightarrow \pi^0 \gamma)=\frac{\alpha g_{\omega\rho\pi}^2}{12}(7/4)^2f_\pi^2m_\omega\left(\frac{m_\omega}{m_\rho}\right)^2 \left(1-\frac{m_\pi^2}{m_\omega^2}\right)^3
\end{equation}
Using the experimental branching ratio for the process $BR(\omega \rightarrow \pi^0 \gamma)=8.28 \pm0.28$ \% \cite{pdg}, we obtain the following value for the individual 
$\omega-\rho-\pi$ coupling:
\begin{equation}
g_{\omega\rho\pi}= 7.7 \pm 0.2  \ {\text GeV}^{-1} \ ({\text from} \ \omega \rightarrow \pi^0 \gamma).
\label{gorp}
\end{equation}

\subsection{The $\omega \rightarrow 3\pi$ decay}
Proceeding along the same lines of the previous subsection, this process is only sensitive to the $\rho$ meson radial excitations. Then, we can take into account their contributions considering the amplitude for the vector channel to be given by:
\begin{eqnarray}
\mathcal{A}^{VMD}_{\omega3\pi}&=&
6\sum_{\rho_n} \frac{g_{ \omega \rho_n \pi } g_{\rho_n\pi\pi}}{m_{\rho_n}^2}
=6 \frac{g_{\omega\rho \pi }g_\rho}{m_\rho ^2} \sum _{n=0}^{\infty } \frac{1}{(1+an)^2}\nonumber\\
&=&(7/4) 6\frac{g_{\omega\rho \pi }g_\rho}{ m_{\rho }^2},
\label{ampres}
\end{eqnarray}
while the purely axial contribution in the contact term remains the same. Thus, requiring this description to account for the experimental decay width of the process, we obtain: \begin{equation}
g_{\omega\rho\pi}= 8. 9 \pm 0.2  \ {\text GeV}^{-1} \ ( {\text from} \ \omega \rightarrow 3\pi),
\end{equation}
which is comparable  with the one obtained from the $\omega \to \pi\gamma$ decay. The difference between them might be considered as a rough estimate of the different role of the model assumptions in each case. In particular, the $\omega \rightarrow \pi^0 \gamma$ decay uses assumption ii) to induce the squared mass dependence while in the $\omega \rightarrow 3\pi$ decay it comes from the propagator, such that a slight modification (about 7\%) on assumption ii) can bring them into agreement. In addition, a neglected momentum dependence, when matching with the experimental width, may also produce a difference of about 6\%. Taking into account all these facts, we set the global average  to be: 
 \begin{equation}
g_{\omega\rho\pi}= 8.3 \pm 0.8  \ {\text GeV}^{-1}.
\label{grho}
\end{equation}

\subsection{The $\pi \rightarrow \gamma\gamma$ decay}
So far we have considered processes where only the $\rho$ radial excitations are present. The $\pi \rightarrow \gamma\gamma$  decay involves not only the $\rho$ but also the $\omega$ excitations. In the following we will compute the contributions from the $\rho$ radial excitations, using the $g_{\omega\rho\pi}$ previously obtained (Eqn. \ref{grho}), and stablish a relation with $g^{eff}_{\omega\rho\pi}$.\\
  In the VMD description, the coefficient in the amplitude, Eqn. (\ref{piggamplitudes}),  including an infinite sum of $\rho$-like contributions, becomes :
\begin{equation}
{\cal A}^{VMD+}_{\pi\gamma\gamma}=2e^2\sum_{\rho_n}
\left( \frac{ g_{\omega{\rho_n}\pi} }{ g_\omega g_{\rho_n}} \right) ,
\end{equation}

Using the assumptions about the couplings and masses of the vector mesons, it takes the following form:
\begin{eqnarray}
2e^2\sum_V\frac{ g_{\omega{\rho_n}\pi} }{ g_\omega g_{\rho_n}} &=& \frac{4e^2 g_{\omega\rho\pi} f_\pi^2}{ 3m_\rho^2}\sum_n \frac{1}{(1+an)^2} \nonumber \\
&=&\frac{4 e^2g_{\omega\rho\pi} f_\pi^2}{ 3m_\rho^2}(7/4).
\end{eqnarray}

The consistency with the anomaly, Eqn. (\ref{piggamplitudes}), requires that:
\begin{eqnarray}
 g_{\omega\rho\pi} &=&\frac{ 3m_\rho^2}{28 \pi^2f_\pi^3}=8.1 \ {\text GeV}^{-1}\nonumber\\
 &=& g^{eff}_{\omega\rho\pi}\frac{4}{7}
\end{eqnarray}
which is consistent with the extracted value for the $g_{\omega\rho\pi}$ coupling, Eqn. (\ref{grho}). Inversely, if we take the value from Eqn. (\ref{grho}), the last relation for the effective coupling gives:
\begin{equation}
g^{eff}_{\omega\rho\pi}= 14.5 \pm 1.4  \ {\text GeV}^{-1},
\label{e2gorp}
\end{equation}
we can observe that this value is consistent with the corresponding one obtained in section II.A  (Eqn. \ref{e1gorp}).
Note that  $g^{eff}_{\omega\rho\pi}$, in that case, was defined as the one that resumed all the vector contributions and made to agree with the anomaly. Thus, the above result tell us that the individual channel coupling is about 60\% smaller than that value.
It is trivial to check that by using this value when adding all the contributions we recover the amplitude of the previous section.

 \subsection{The $\gamma^* \rightarrow 3\pi$ decay}
The $\gamma^* \rightarrow 3\pi$ decay is similar to the $\pi \rightarrow \gamma\gamma$ in the sense that not only the $\rho$ but also the $\omega$ excitations are involved. However,
as we have seen previously,  it is also affected by the  contact term. Let us see the individual contributions upon the inclusion of the $\rho$ radial excitations. The amplitude for the $\rho$-like channel  is given by:
\begin{eqnarray}
\mathcal{A}^{VMD+}_{\gamma3\pi}&=&
\frac{6e}{g_\omega} \sum _{\rho_n} \frac{g_{\omega {\rho_n} \pi } g_{{\rho_n}\pi\pi}}{m_{\rho_n}^2}
=2 e \frac{g_{\omega\rho \pi }}{m_\rho ^2} \sum _{n=0}^{\infty } \frac{1}{(1+an)^2}\nonumber\\
&=&2e \frac{g_{\omega\rho \pi }}{ m_{\rho }^2}(7/4)=\frac{3}{2} \mathcal{A}^{WZW}_{\gamma3\pi}
\label{ampres}
\end{eqnarray}
That is, by adding all the vector contributions we recover the result of the previous section for the $\rho$, but now including all its radial contributions.

The purely axial contribution in the contact term remains the same:
\begin{equation}
\mathcal{A}^c=\frac{-1}{2}\mathcal{A}^{WZW}_{\gamma3\pi},
\end{equation}
\noindent while the relationship for the $\rho$ channel ($n=0$) now accounts for 
\begin{equation}
\mathcal{A}^\rho_{\gamma3\pi}=2e \frac{g_{\omega\rho\pi }}{m_\rho ^2} =\frac{3e}{8\pi^2 f_\pi^3} \frac{4}{7 }=  \left(\frac{4}{7 }\right)\frac{3}{2} \mathcal{A}^{WZW}_{\gamma3\pi}.
\end{equation}

Since we have estimated all the vector contributions, the radial excitations contributions are just that with the $\rho$ contribution removed:
\begin{equation}
\mathcal{A}^{re}_{\gamma3\pi}=2e \frac{g_{\omega\rho \pi }}{m_{\rho }^2}(\frac{7}{4}-1).
\end{equation}

We can also treat this contribution as an effective contact diagram and estimate a magnitude for its effective coupling $g^{re}_{3\pi}$. 

\begin{equation}
\frac{6 eg^{re}_{3\pi}}{g_{\omega }}=2e \frac{g_{\omega\rho \pi }}{m_{\rho }^2}(\frac{7}{4}-1),
\end{equation}	
from this equivalence we get 
\begin{equation}
g^{re}_{3\pi}= \frac{g_{\omega\rho\pi}}{\sqrt{2}m_\rho f_\pi}(\frac{7}{4}-1)=61\pm 6 \ {\text GeV}^{-3},
\end{equation}
where we have used the value of $g_{\omega\rho\pi}$ as in Eqn. \ref{grho}.
Therefore, the radial excitations and axial contributions combine to account for a fraction of the total amplitude given by:
  \begin{equation}
  \mathcal{A}^{re} + \mathcal{A}^{c} =\frac{1}{7} \mathcal{A}^{WZW}_{\gamma3\pi}.
  \end{equation}
 Note that the global factor suppresses this contribution, and justify the observation that the single $\rho$ channel accounts for most of the total decay.\\

\begin{table}
\begin{center}
\begin{tabular}{|c |c|c|}
    \hline
     Coupling & Process & Value\\
     \hline \hline 
      $|g_{\omega \rho \pi}|$ &$\omega \rightarrow \pi^0 \gamma$& 7.7 $\pm$ 0.2 GeV$^{-1}$ \\  
                                          &$\omega \rightarrow 3\pi$& 8.9 $\pm$ 0.2 GeV$^{-1}$ \\ \hline 
      $|g^{eff}_{\omega \rho \pi}|$ & $\pi^0 \rightarrow \gamma \gamma$ &14.2 GeV$^{-1}$ \\
       &$\omega \rightarrow 3\pi$& 12.8  GeV$^{-1}$ \\

       \hline 
      $g^c_{3\pi}$&$\gamma^* \rightarrow 3\pi$ & -47 GeV$^{-3}$\\ \hline
      $g^{re}_{3\pi}$&$\gamma^* \rightarrow 3\pi$ & 61$\pm$ 6 GeV$^{-3}$ \\ \hline
    \hline
\end{tabular}
\end{center}
\caption{Couplings obtained from a set of processes.}
\label{gwrp}
\end{table}

In table I we summarize the numerical results for the couplings in the different scenarios we have considered.\\

\section{Discussion}
The extraction of the $g_{\omega\rho\pi}$ coupling is unavoidable made by indirect means. In this work we have pointed out that the usually quoted value corresponds not to it but to another $g^{eff}_{\omega\rho\pi}$ effective coupling  which resumes not only the $\rho$ meson effect but also all its radial excitation modes.\\
Here, we have explicitly added the radial excitations of the $\rho$ meson, considering a particular form of the spectrum and relations among the couplings. First, we considered the $\omega \rightarrow \pi^0 \gamma$ and  $\omega \rightarrow 3\pi$ decays to identify the single  $g_{\omega \rho \pi}$ and the $\rho$ radial excitations effect. This process is clean in the sense that it is sensitive to the $\rho$ radial excitations but not to the $\omega$ ones. This fact allows to avoid any assumption on the $\omega$ excitations.
Certainly,  the description used for all those excitations is model dependent. However, the lowest lying excitation is well approached and expected to be the dominant one. Besides this model dependence, our description succeeds in exhibiting how each contribution came into the game while fulfilling general requirements like the agreement between the VMD and chiral anomaly descriptions. 
We obtained that the individual coupling is $g_{\omega\rho\pi}= 8.3 \pm 0.8 \ {\text GeV}^{-1}$, which is about 40\% smaller than the effective  $g^{eff}_{\omega\rho\pi}$.\\
We have verified the consistency with the chiral approach in the  $\pi^0 \rightarrow \gamma\gamma$ and $\gamma^* \rightarrow 3\pi$ processes. Our description succeeds in exhibiting how each contribution came into the game. In particular,  we show that, for the $\gamma^* \rightarrow 3 \pi$ decay, the usual relation $\mathcal{A}^{VMD+}_{\gamma3\pi}=(3/2)\mathcal{A}^{WZW}_{\gamma3\pi}$, encodes all the vector contributions and not only the $\rho$ meson one. Thus, the additional contact term is fully axial and fixed by the WZW anomaly. In addition, we have obtained that there is a strong cancelation between the radial excitations and the contact term contributions.\\
The relations here stablished are hold in the zero momentum limit. More elaborated assumptions would be required to explore in a reliable way the momentum dependence.\\
We would like to conclude by stressing that the usually neglected contributions from radial excitations  may be relevant  even though  they can be very heavy and must be carefully considered.

\begin{acknowledgments}
We acknowledge the support of CONACyT-M\'exico under grant 128534, and DGAPA-UNAM under grants PAPIIT-IB101012 and PAPIIT-IN106913 .
\end{acknowledgments}

\end{document}